\documentclass[conference]{IEEEtran}
\IEEEoverridecommandlockouts
\usepackage{cite}
\usepackage{amsmath,amssymb,amsfonts}
\usepackage{algorithmic}
\usepackage{graphicx}
\usepackage{textcomp}
\usepackage{xcolor}
\def\BibTeX{{\rm B\kern-.05em{\sc i\kern-.025em b}\kern-.08em
    T\kern-.1667em\lower.7ex\hbox{E}\kern-.125emX}}

\usepackage{balance}
\IEEEpubid{\begin{minipage}{\textwidth}\ \\[12pt]
978-1-7281-1884-0/19/\$31.00 \copyright 2019 IEEE
\end{minipage}}

\begin{document}

\title{Enhancing Battle Maps through Flow Graphs}

\author{\IEEEauthorblockN{G\"unter Wallner}
\IEEEauthorblockA{\textit{Institute of Art \& Technology} \\
\textit{University of Applied Arts Vienna}\\
Vienna, Austria \\
guenter.wallner@uni-ak.ac.at}
}
%In this paper we propose an additional step to the original battle map algorithm proposed by Wallner~\cite{Wallner:2018}

\maketitle

\begin{abstract}
So-called battle maps are an appropriate way to visually summarize the flow of battles as they happen in many team-based combat games. Such maps can be a valuable tool for retrospective analysis of battles for the purpose of training or for providing a summary representation for spectators. In this paper an extension to the battle map algorithm previously proposed by the author~\cite{Wallner:2018} and which addresses a shortcoming in the depiction of troop movements is described. The extension does not require alteration of the original algorithm and can easily be added as an intermediate step before rendering. The extension is illustrated using gameplay data from the team-based multiplayer game \emph{World of Tanks}.
\end{abstract}


\section{Introduction}
	\label{sec:introduction}
		
Information visualization has become an indispensable element for analyzing behavioral data of players as reflected in the increasing efforts and literature on this topic (see~\cite{Wallner:2013}). In this context, visualization is usually considered a means for developers to analyze the collected data in order to inform the further development or to drive business decisions. 

However, visualizations can also be targeted towards players. Visualizations within games historically served mainly to convey the in-game status to players~\cite{Bowman:2012}. Recently, fueled by developers making in-game data publicly available, visualizations for the purpose of training with the goal to improve players' skills and performance (cf.~\cite{Bowman:2012,Hazzard:2014}) have also gained in popularity. Moreover, visualizations can also benefit spectators of esports events -- on-site or, similar to traditional sport events~\cite{Pingali:2001}, when streamed or broadcast. Such training visualizations and visualizations for enhancing the spectator experience of video games have, however, received comparatively less attention among academia despite visualizations being part of games since the very early days. 

Notable examples in the space of esports include the work of Block et al.~\cite{Block:2018} who generated audience-facing summary visualizations from live and historic match data and from Charleer et al.~\cite{Charleer:2018} who proposed dashboards superimposed over the game stream and which show live statistics about the current match status. Common to both works is their focus on presenting data about the current state of the game rather than a synthesized summary of the complete match. Training visualizations, on the other hand, are mostly produced by the player community itself, creating tools such as \emph{Scelight}~\cite{Scelight} for analyzing \emph{StarCraft II} matches. 

Among video games, team-based games -- as team sports -- pose additional requirements on visualization as these need to reflect the collaborative behaviour of multiple players (cf.~\cite{Page:2006}). At the same time, players and viewers of such games can benefit strongly from visualizations that facilitate the understanding of player activity and their coordination. Wallner and Kriglstein~\cite{Wallner:2016} investigated three different summary representations for retrospective analysis of team-based combat games for the purpose of training, including battle maps. These maps, inspired from depictions of battles by historians and military planers, provide a concise overview of unit movements and encounters.

This short paper reports on new progress with respect to automatically creating battle maps from in-game data by extending the algorithm proposed in previous work of the author~\cite{Wallner:2018} to further enhance their readability. This is accomplished by merging individual troop movements to better highlight the splitting and merging of troops and their total strength.

\IEEEpubidadjcol

%  Player-centric visualizations can, however, also serve the purpose of training with the goal to improve the players’ skills and performance (cf.~\cite{Hazzard:2014}).

\section{Related Work}
	\label{sec:relatedwork}
	
Aggregation of movement data takes on a crucial role in geographic and traffic data visualization (e.g.,~\cite{Andrienko:2008,Chen:2015}). While approaches in these domains are certainly relevant for the aggregation of player movements, they are too numerous to cover here in detail. We will therefore restrict the discussion to works specifically conducted within games research. 

Examples in this space include the work of Wallner et al.~\cite{Wallner:2019} who also made use of the same territory subdivision technique~\cite{Adrienko:2011} as the battle map algorithm~\cite{Wallner:2018} extended in this work. However, instead of using the derived cells to infer abstracted trajectories the cells are directly used to show the amount of movement between them. Moura et al.~\cite{Moura:2011} also showed player flows between areas but instead of deriving the areas from the movement data itself the areas are defined based on the level geometry (e.g., rooms). Canossa et al.~\cite{Alessandro:2016}, on the other hand, make use of heatmaps to convey in which regions movement has occurred, thus showing rather the amount of movement that occurred without indication of the direction. 

More closely related to our work, Miller and Crowcroft~\cite{Miller:2010} used similar techniques to detect group movements in \emph{World of Warcraft}. Avatar movement was characterized using hotspots which, in turn, where detected by subdividing the environment into regular cells and assessing how much time the avatars spend in these cells (hotspots are also part of the battle map algorithm, but these are derived through the clustering of combat points). Two avatars that move between two hotspots and maintaining a certain distance are then considered to move together. Affiliation to the same group is thus defined through the geospatial locations while the battle map defines it through the similarity of semantic trajectories. 

\section{Algorithm}
	\label{sec:algorithm}
	
\begin{figure*}[t]
    \centering
    \includegraphics[width=0.445\linewidth]{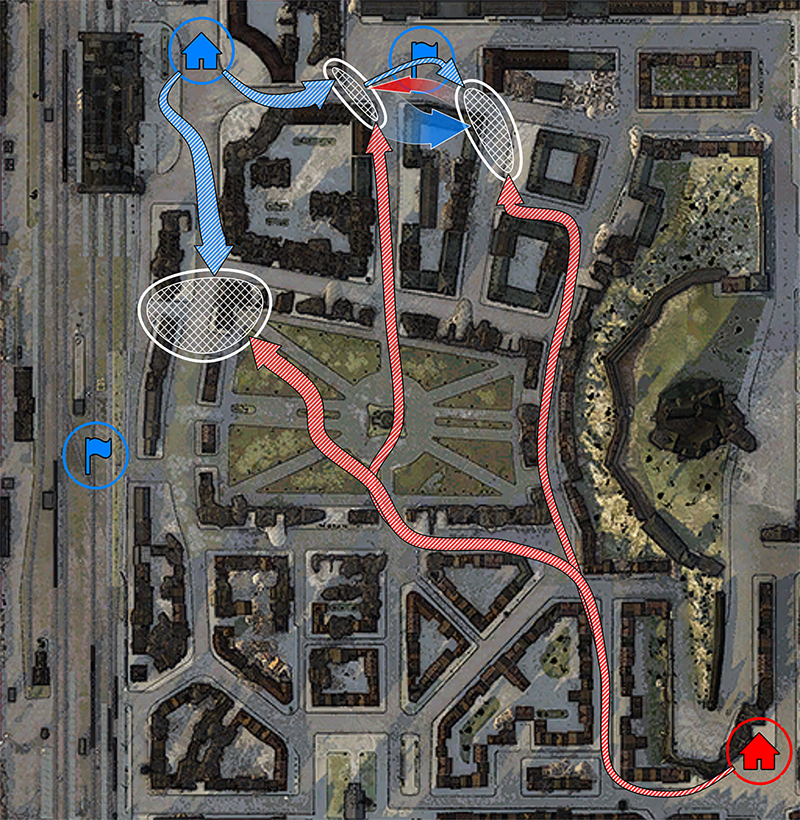}\hskip5pt
    \includegraphics[width=0.445\linewidth]{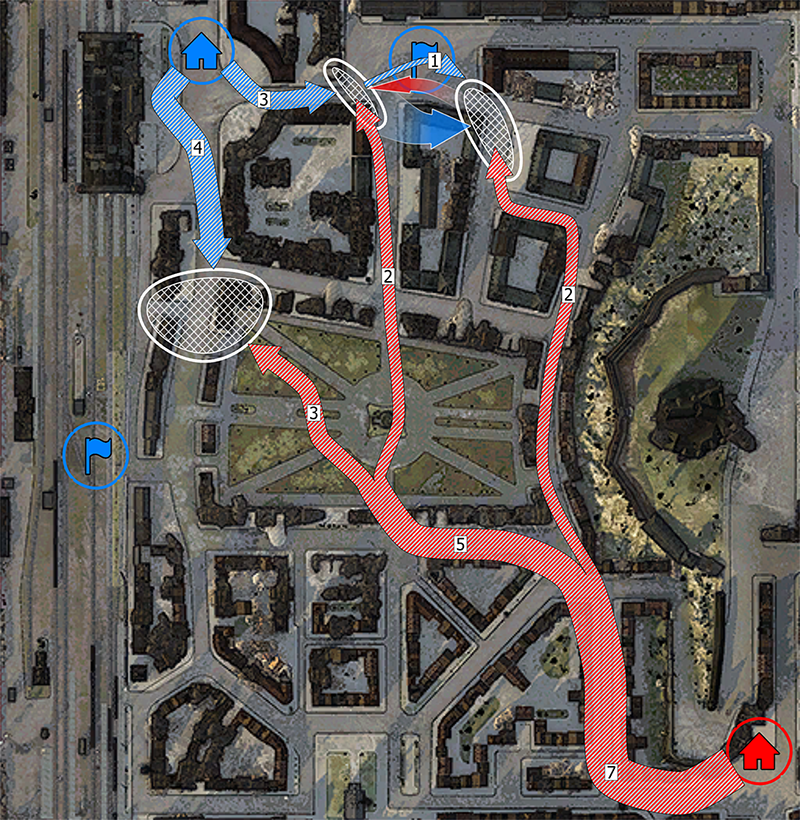}
    \vskip4pt
    \includegraphics[width=0.445\linewidth]{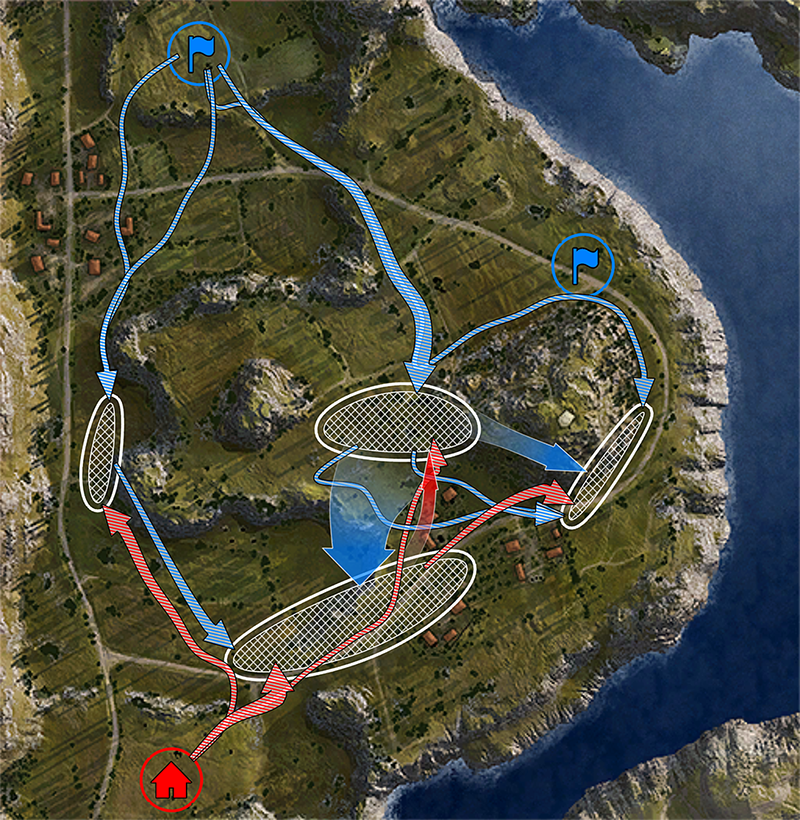}\hskip5pt
    \includegraphics[width=0.445\linewidth]{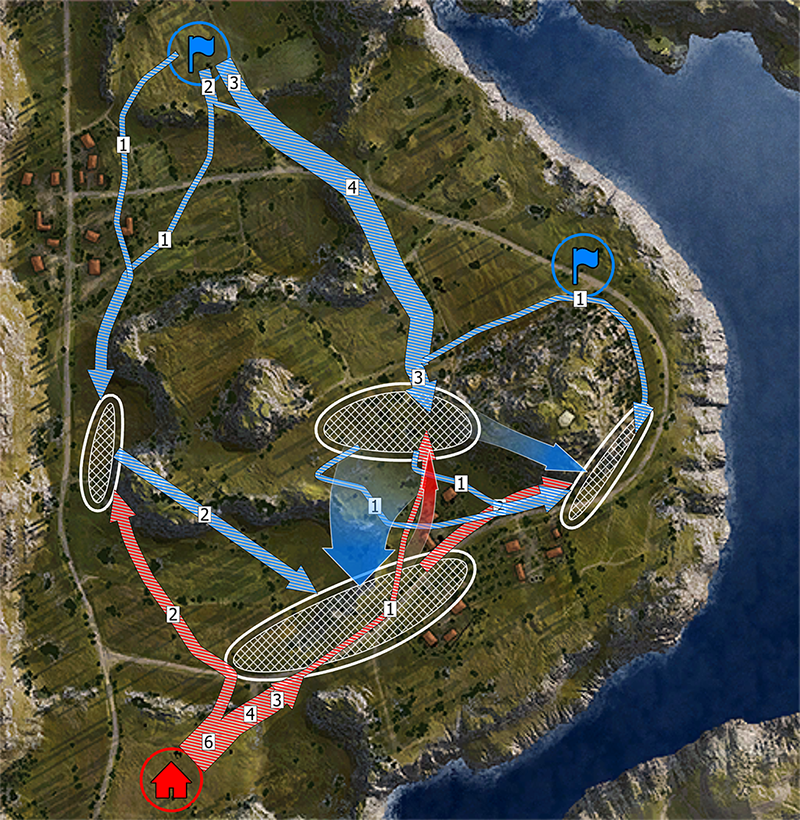}
    \caption{Comparison of resulting battle maps without (left) and with the proposed extension (right) based on two matches played on the maps \emph{Himmelsdorf} (top) and \emph{Cliff} (bottom) of \emph{World of Tanks}.}
    \label{fig:comparison}
\end{figure*}

In previous work, Wallner~\cite{Wallner:2018} proposed an algorithm for automatically deriving battle maps from tracked gameplay data. It can be briefly summarized as follows: First, the territory of the game map is subdivided into small cells. Next, trajectories of all involved units are simplified by replacing them with semantic trajectories, that is, sequences of visited locations (i.e. landmarks) based on the recorded geospatial positions and the derived cells. These semantic trajectories are then partitioned into sets according to their origin and destination, with trajectories within each set being further grouped based on their similarity to each other (since the same destination can be reached from the same origin by taking different routes). For each group a representative trajectory is calculated that represents the overall movement of the group. Finally, these representative trajectories are used for visualizing the group movements. 

Resulting battle maps obtained using this algorithm and by processing replay files from the game \emph{World of Tanks}~\cite{wot:pc} are depicted in Figure~\ref{fig:comparison} (left). Shortly summarized, \emph{World of Tanks} is an online warfare game where two teams of players compete against each other for certain objectives, with each player controlling a single tank. Once a player's tank is destroyed the player cannot re-enter the match. The hatched arrows in Figure~\ref{fig:comparison} show troop movements, while gradient shaded arrows show long distance attacks. Sites of combat are enclosed within a white curve and bases and spawn points are shown using icons.

While an evaluation~\cite{Wallner:2016} attested the visualization on overall good readability it also revealed that judging troop movements correctly can be prone to errors, most likely due to the overlapping movement arrows. For example, in Figure~\ref{fig:comparison} (top, left) the troops of the red team leaving the base in the lower right corner all move along the same path and are thus occluded until they branch off. This makes judging the actual amount of troops during the first segments of the path more difficult. The increased width of the arrows with accumulated travel distance may also affect the perception of troop strengths. This paper addresses these issues by proposing an intermediate step before rendering as detailed in the following. As no adjustments to the steps before need to be made it can be easily integrated into the original algorithm. 

\subsection{Extension}
	\label{sec:extension}

The basic idea is to replace the individual representative trajectories with a graph structure in order to reduce occlusions and better show the splitting and merging of troops. The approach is similar to that of a flow tree~\cite{Phan:2005} which connects a single origin with multiple destinations. However, since the merging of troops should also be visible we use a graph structure (henceforth referred to as flow graph) -- more specifically a weighted directed acyclic graph -- instead of a tree. It should be observed that the graph is acyclic because it summarizes unidirectional movement from one source to multiple destinations. If troops would, for example, backtrack this would be captured by another flow graph. Also, these flow graphs are constructed separately for each team and for each origin.

Given a number of representative trajectories all starting at the same origin $o$ and represented by a sequence of landmarks $o=l_1 \rightarrow l_2 \rightarrow l_3  \rightarrow \ldots \rightarrow l_k$ with each landmark associated with a location $p_i$, first all landmarks within the enclosure of the last landmark $l_k$ (i.e. destination) are removed from the sequence. Next all transitions between landmarks $l_i \rightarrow l_{i+1}$, are added as directed edges $(l_i, l_{i+1})$ to a directed acyclic graph $G$ where the nodes represent the landmarks occurring in the representative trajectories. If an edge already exists than the weight of the edge is incremented. This results in a graph summarizing the transitions between landmarks. 

Next, to ensure a smooth representation of the flow of movement, a cubic Hermite spline is derived for each edge $(l_0, l_1)$ in a flow graph starting from its root node. The tangent $t_0$ at the starting position of the spline is set to the principal movement direction at $l_0$, that is, the average direction of all incoming and outgoing movement at $l_0$. This way all the outgoing flow at a node will leave it along the same direction. If troops split up at the node than the curve starts in the 'middle' of the outgoing directions while also accounting for the directions along which the units arrived at a particular node. Similarly, the tangent at the end position is set to the average direction at $l_1$. Note, that this ensures $C_1$ continuity. 

To avoid that splines originating from the same node would all start at the same location, the outgoing edges of $l_0$ are sorted clockwise around $p_0$ and the starting points of the corresponding splines are displaced along the normal of $t_0$ based on the edge weights. Similarly, the incoming edges at $l_1$ are sorted clockwise around $p_1$ and the ending points of the associated splines are offset along the normal of $t_1$. This avoids occlusions of the splines at the nodes and reduces crossings of the outgoing and incoming flow. 

Once all edges are processed, this process yields a piecewise cubic spline representation of the flow graph which can be used for rendering it in a visually appealing way. The width (normalized with respect to the maximum troop size) reflects the number of units but in contrast to~\cite{Wallner:2016} is kept constant over traveled distance do not make the impression that the troop strength changes.

Lastly, labels are added to offer exact values on the number of units. However, as the flow does not branch at each node but rather only at a few nodes, labels are not added to each edge (i.e. spline) but instead only to segments between two nodes where each node has more than one outgoing edge. 

\section{Results and Discussion}
	\label{sec:results}

%	\caption{Comparison of resulting battle maps without (left) and with proposed extension (right) accentuating total troop strength and how they split up.}
%  Reuniting troops (blue team) and combined troop strength are visually emphasized.

Figure~\ref{fig:comparison} compares results obtained without (left) and with (right) the extension by the example of two matches played on two different maps of \emph{World of Tanks}~\cite{wot:pc}. All other parameters of the algorithm such as the similarity of paths required to group them together were kept constant. 

The images on the top show a battle fought on the map \emph{Himmelsdorf}, a confined urban map composed of squares and narrow streets. The most visible difference is with respect to the red team where without the extension the arrows of troops leaving the base at the lower right corner overlap. As such the number of units might be perceived to be lower than it is actually the case. In contrast, with the extension the width of the arrow better accentuates the total troop strength (e.g., immediately after leaving the base) and also how the troops split up. It is thus likely to give a better impression of troop strength over the course of movement.

In contrast, Figure~\ref{fig:comparison} (bottom) shows a match played on \emph{Cliff}, a more open map which allows for more variation in unit movement, with troops splitting up and then reuniting again. Since troops after merging at a location -- as it is the case twice for the blue team -- are represented by a single but thicker arrow, the merging and the resulting troop strength are visually emphasized compared to the result without the extension. 

It should be noted that the time dimension of troop movements needs to be taken into account when merging individual representative trajectories into a flow graph. Merging movements which took place at very different points in time would create a false impression of the actual tactical movements of units. In the above examples, this was implicitly considered as the representative trajectories reflect movements which originated and ended at a location within a certain time span. Another approach could be to only merge movements if the units where present at roughly the same time at the semantic area of interest. Future work may also focus on splitting activity into time slots to produce multi-stage battle maps which could improve perception of the time dimension further. Finally, follow up work will need to focus on evaluating the extension to assess if the proposed changes lead to the anticipated improvement in readability.

\section{Conclusions}
	\label{sec:conclusions}
	
This paper proposed a way to merge individual troop movements into combined flows in order to improve the readability of troop movements, their splitting and merging, and their overall strength in the context of battle maps. Such maps have shown to be a valuable tool for players to reflect on their gameplay~\cite{Wallner:2016}. In addition, they could also be a promising asset in the context of esports. For example, by offering a summary visualization for the viewers or as an aid for shoutcasters. While this paper focused on the algorithmic aspects, further work will need to evaluate the proposed technique to ascertain the expected improvement. 

\balance
\bibliographystyle{IEEEtran}
\bibliography{references}

\end{document}